# Indirect to direct bandgap transition in methylammonium lead halide perovskite

**Authors:** Tianyi Wang[1†], Benjamin Daiber[1†], Jarvist M. Frost[2], Sander A. Mann[1], Erik C. Garnett[1], Aron Walsh[2], Bruno Ehrler[1]*

**Affiliations:**

[1] Center for Nanophotonics, FOM Institute AMOLF, Science Park 104, 1098 XG Amsterdam, The Netherlands.

[2] Centre for Sustainable Chemical Technologies and Department of Chemistry, University of Bath, Claverton Down, Bath BA2 7AY, United Kingdom.

*Correspondence to: ehrler@amolf.nl

†These authors contributed equally to the work.

**Abstract**

Methylammonium lead iodide perovskites are considered direct bandgap semiconductors. Here we show that in fact they present a weakly indirect bandgap 60 meV below the direct bandgap transition. This is a consequence of spin-orbit coupling resulting in Rashba-splitting of the conduction band. The indirect nature of the bandgap explains the apparent contradiction of strong absorption and long charge carrier lifetime. Under hydrostatic pressure from ambient to 325 MPa, Rashba splitting is reduced due to a pressure induced ordering of the crystal structure. The nature of the bandgap becomes increasingly more direct, resulting in five times faster charge carrier recombination, and a doubling of the radiative efficiency. At hydrostatic pressures above 325 MPa, MAPI undergoes a reversible phase transition resulting in a purely direct bandgap semiconductor. The pressure-induced changes suggest epitaxial and synthetic routes to higher efficiency optoelectronic devices.

**Introduction**

Solar cells based on methylammonium lead iodide perovskites (MAPI) have seen an unprecedented increase in efficiency over a short period of time[1-4], while other applications including lasers and photodetectors have shown promise[5,6]. The high efficiency of these applications arises due to low defect density[7] with long charge carrier lifetime[8] and diffusion length[9], in spite of the material being solution processable. To date, MAPI has widely been considered a direct bandgap semiconductor according to both theoretical calculations and experimental observations[10,11]. However, the unusually long minority carrier lifetime with values more similar to those of indirect bandgap semiconductors[12], and the associated long charge carrier diffusion length has been a long-standing mystery in the field. It evoked explanations based on long-lived trapping of charges[13], large polarons[14], and triplet exciton formation[15]. Recently theoretical calculations predicted a slightly indirect bandgap in this material[16]. Brivio et al. calculated the band structure of MAPI using quasiparticle self-consistent *GW* theory and found that a Rashba-splitting of the conduction band should generate this slightly indirect bandgap[17]. The same relativistic effect has been reported in other calculations[18-20].

Recently, first experimental indications point towards a Rashba-split band in $MAPbBr_3$[21]. However, there is no direct experimental evidence that supports the theoretical predictions of the bandstructure of the prototypical solar cell material MAPI, or the dramatic consequences for the charge carrier dynamics. The bandstructure of a semiconductor can be altered by structural changes under external application of pressure[22,23]. Pressure has been applied to MAPI to understand the structural changes[24-26]. Recently, it was also used to study the recombination dynamics[27]. It was found that MAPI undergoes a phase transition at around 325 MPa The phase at ambient pressures is tetragonal[23,25] and the high pressure phase has been subject to debate, assigned to orthorhombic[24,26,27] and cubic[25] crystal phases. A further phase transition is known to occur in the GPa regime[24-26].

Here we find that MAPI has an indirect bandgap 60 meV below the direct bandgap both in absorption and emission spectra. This indirect gap is responsible for the unusually long carrier lifetime because the thermalized carriers are protected against recombination *via* the fast direct transition. The indirect transition arises from Rashba splitting of the conduction band. The band is split due to the local electric field generated by the absence of inversion symmetry around the Pb site, which acts on the 6p orbitals of the lead-atom where most of the conduction band minimum is located[17,28].

We study the optoelectronic changes of thin polycrystalline MAPI films under mild hydrostatic pressure up to 400 MPa, below and just above the phase transition at 325 MPa. We show that the bandgap changes with pressure and that the direct transition is enhanced. Above the phase transition MAPI behaves like a purely direct bandgap semiconductor. As the bandgap becomes more direct, the charge carrier lifetime decreases drastically with increasing pressure and the photoluminescence quantum efficiency (PLQE) exhibits a two-fold increase. These changes can be understood in terms of an increase in inversion symmetry around the Pb site under pressure and thus a reduction in Rashba splitting, resulting in a more direct bandgap of MAPI. Our results show that small structural changes can significantly improve relevant optoelectronic properties of MAPI.

**Results and Discussion**

We apply hydrostatic pressure to 400 nm thin polycrystalline films of MAPI using a pressure cell filled with an inert, mechanically pumped pressure liquid (see Methods for details). The material is continuously compressed with a maximum strain of 3% at 400 MPa, as calculated from a Young's modulus of 12.8 GPa[29]. This value is in good agreement with the change of lattice volume up to 400 MPa, derived from powder x-ray diffraction (PXRD) data[25]. The absorption spectrum remains constant in shape, but the onset of absorption clearly changes with pressure (Error! Reference source not found.**a**, and **Fig. S1** for complete set of absorption spectra). We used the linear part of Tauc plots[30] of the absorption edge to extract the (direct) bandgap (see **Supplementary Information S1** for details). For the pressure upstroke, a continuous red-shift of the absorption edge by 30 meV is observed until 325 MPa. This unusual trend arises due to the electronic band structure associated with the corner sharing $PbI_3^-$ network[31]. While many semiconductors have negative bandgap deformation potentials (the bandgap increases with increasing pressure), hybrid perovskites have positive values as the band extrema are found at the boundary of the Brillouin zone ($R$ point) that corresponds to out-of-phase interactions between neighboring unit cells[32].

An abrupt blue-shift occurs above 325 MPa which can be understood as a result of a phase transition from a tetragonal to an orthorhombic or cubic crystal structure[24-27]. The phase transition is followed by a continuous blue-shift. Upon release of pressure from 400 MPa, i.e. the pressure downstroke, the change in bandgap is completely reversible, thus we can infer that the deformation of the crystal lattice is elastic and therefore the pressure-induced structural changes, including the phase transition, are reversible. The change in bandgap is in good agreement with the result of density functional theory (DFT) calculations before the phase transition (dashed line in Error! Reference source not found.**b**, see **Experimental methods** for details). Atomistic modeling was not attempted on the high-pressure structure that emerges above 325 MPa, so the secondary blue-shift is not recovered in the calculations.

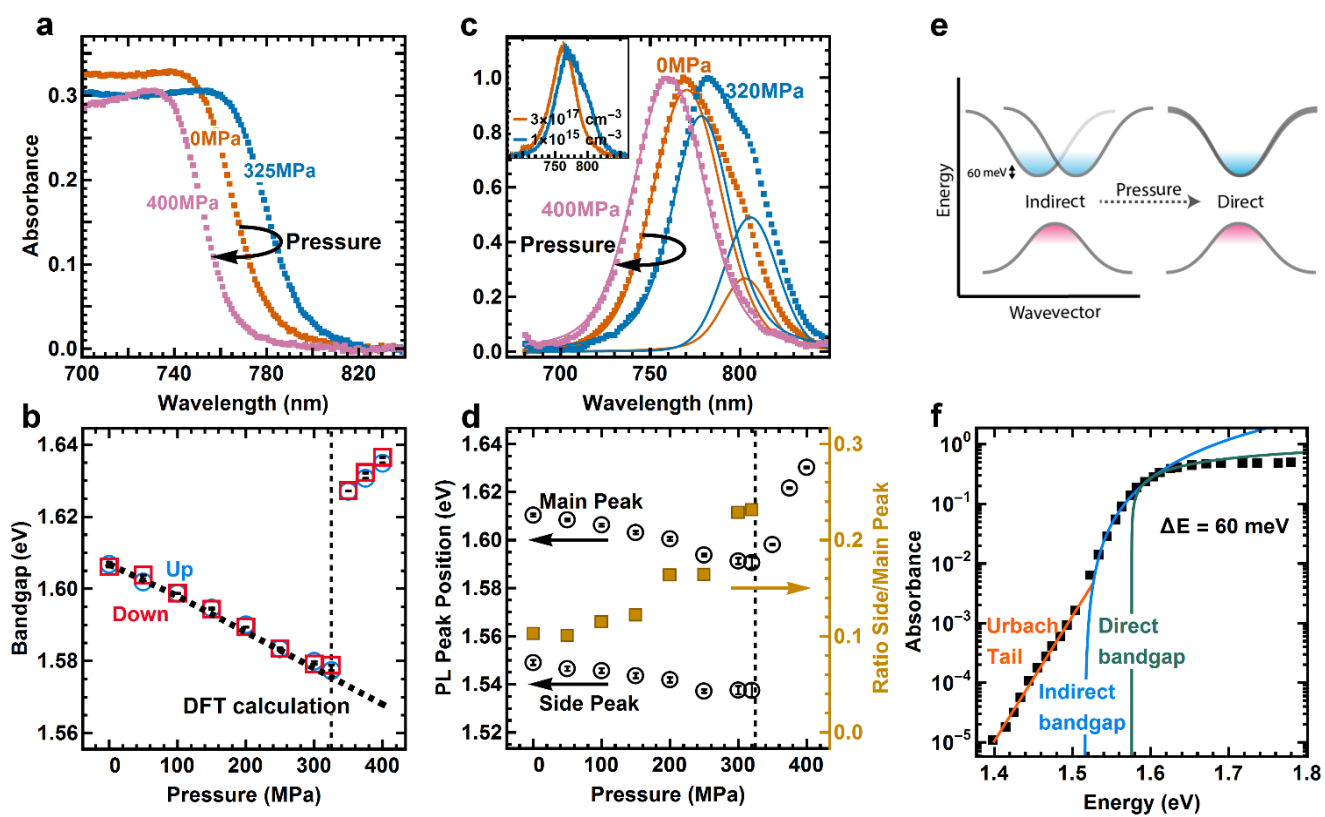

**Fig. 1** Absorbance and steady-state photoluminescence (PL) of MAPI at different pressures. (a) Absorbance spectra at three characteristic pressures. (b) Change of the direct bandgap under pressure for pressure up (blue circles) and down (red squares) stroke. DFT calculations (dashed line) predict the bandgap change before phase transition. Black bars inside symbols represent the error from the fit. (c) Photoluminescence (PL) spectra and fits at three different pressures. At pressures below phase transition (325 MPa) the PL fit requires two peaks while above the phase transition only one peak suffices. The inset shows PL spectra at different excitation densities under ambient pressure (d) Position of main and side peak as extracted from fit to PL data. The ratio of main to side peak intensity decreases with rising pressure. (e) Schematics of band structure, showing direct and indirect transitions and their change with pressure. (f) PDS data from literature[39] fit with an exponential Urbach tail (orange) and indirect (blue) and direct (green) bandgap.

The steady-state photoluminescence (PL) spectra of MAPI reveal peculiar behavior under pressure (see Error! Reference source not found.**c**, **Fig. S2** for complete set of PL spectra, and **Supplementary Information S2** for details). We find that the PL spectrum is composed of two distinct peaks before the phase transition, and that one peak is sufficient to fit the data after the phase transition. At 0 MPa, the dominating emission peak is located at around 1.61 eV, while a side emission peak is located at around 1.55 eV. The side peak disappears at high excitation densities (as shown in **Fig. 1c**). Many reports at high excitation densities hence do not resolve the side peak[15,33,34], or only show a slightly asymmetric PL spectrum[27,35]. The main peak has been assigned to band-to-band recombination and speculations about the origin of the side peak include phonon-assisted recombination[34] and exciton-phonon interaction[36]. A PL side peak has also been observed in the low-temperature orthorhombic phase of MAPI. However, that side peak has been assigned to trapped[37,38] or bound[5,15] charge-carrier pairs, which, as we show, is different compared to the side peak reported in this paper. With increasing pressure up to 325 MPa, the main peak exhibits a red-shift by 10 meV and side peak exhibit a red-shift by 20 meV (Error! Reference source not found.**d**). The main peak shows an abrupt blue-shift at the phase transition followed up by a continuous blue-shift up to 400 MPa, which is consistent with the shift of absorption edge. In the following we will show that the side peak arises from an indirect transition 60 meV below the direct bandgap.

Recent theoretical studies have predicted that a Rashba splitting of the conduction band is present for MAPI, leading to a slightly indirect transition 50 to 75 meV below the direct bandgap[28] (Error! Reference source not found.**e**). This is due to strong spin-orbit coupling (SOC) in MAPI crystal. The conduction band in MAPI is mainly formed by the lead (Z=82) 6p orbitals, while the valence band is mainly formed by iodine (Z=53) 5p orbitals. As the spin-splitting scales approximately with $Z^2$, the valence band is less affected than the conduction band. Therefore, although a small splitting of the valence band is observed in the calculations, it is negligible compared to the large splitting in conduction band. The flattened valence band provides a large density of states that enables strong direct absorption. However, when MAPI is excited at low intensities, mainly the bottom of the conduction band will be filled, which means that (slow) indirect transitions will play a significant role.

In our PL spectra, the energy of the side peak is 60 meV below the main peak, which fits the theoretical prediction (the energy of the phonon required is small in comparison, see **Supplementary Information S3** for details). The ratio between the area of main and side peak is decreasing with increasing pressure up to 325 MPa (Error! Reference source not found.**d**). This trend can be seen as a result of pressure-induced ordering that reduces the lead iodide cage distortion (and occupied volume) and hence the electric field at the lead atom. A reduced electric field, acting on the spin-orbit coupling, leads to a reduction in the Rashba splitting of the conduction band. Thus, a smaller change in crystal momentum is required for the indirect transition, rendering it more efficient with a larger partial density of states of acoustic phonons. At the same time the reduced splitting of the conduction band moves the indirect transition closer in energy to the direct transition, from 60 meV to around 50 meV at 300 MPa (**Fig. S3**). After the phase transition at 325 MPa only one PL peak is present, indicating that MAPI has transformed into a purely direct bandgap semiconductor. The indirect bandgap is also apparent in the absorption spectrum below the direct bandgap. We find that data gathered from photothermal deflection spectroscopy (data taken from literature[39]) can only be fitted with both a direct bandgap and an indirect bandgap between 46 and 60 meV below the direct gap, in addition to the Urbach tail (one example in **Fig. 1f**, see **Supplementary Information S4** for details and **Fig. S8-S10** for additional spectra). Combining only a direct bandgap and the Urbach tail (dashed line in **Fig. S8-S10**) yields a very poor fit.

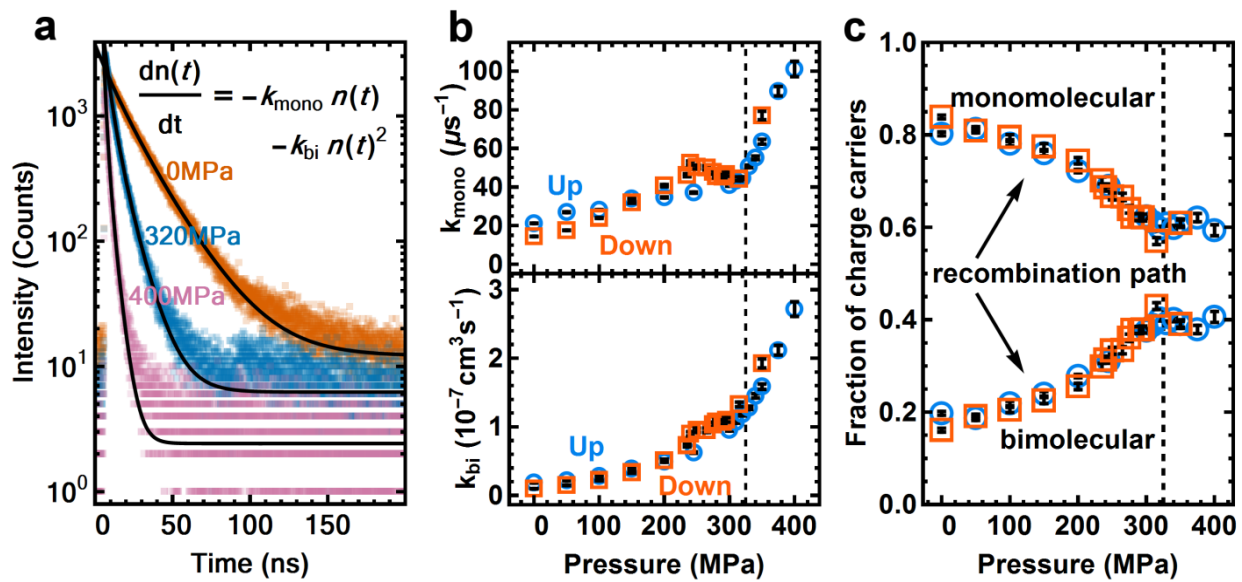

**Fig. 2** Time-Correlated Single Photon Counting (TCSPC) data of the PL of a MAPI film under pressure, excited with a 640nm laser. (A) TCSPC data of 0 MPa, 320 MPa and 400 MPa pressure. The black line shows the fit from the model shown in the inset. (B) Monomolecular (top) and bimolecular (bottom) decay rates of the charge carriers, extracted from the fit. The bimolecular recombination rate (bottom) changes faster with pressure than the monomolecular recombination rate, with both changing slope after the phase transition at 325 MPa (dashed line). Black bars inside symbols represent the error from the fit. (C) Fraction of charge carriers that decayed through the monomolecular channel and bimolecular channel.

We expect the transition towards a more direct bandgap under pressure to result in shorter excited state lifetime. We use time-correlated single photon counting (TCSPC) to probe excited state lifetime at different pressures. The sample is excited with a red laser at 640nm (12 nJ/cm$^2$), at a low initial charge carrier density of around $6\times10^{14}$ cm$^{-3}$ (see **Supplementary Information S5** for details). This mild excitation density reduces the filling of the conduction band and thus enhances the contribution seen from the indirect transition. At a charge carrier density above $10^{17}$ cm$^{-3}$, where bimolecular and monomolecular decay rates increase at higher pressure (**Fig. 2b**). We note that this is contrary to the trend recently measured for MAPI powders[27]. To determine how well bimolecular recombination competes with monomolecular recombination we calculate the fraction of charge carriers that decayed via the bimolecular (radiative) and monomolecular (non-radiative) path (**Fig. 2c**) (see **Supplementary Information S5** for details of calculation) path. The internal radiative efficiency increases from 20 % at ambient pressure to 40 % at 325 MPa and remains at around 40 % from 325 to 400 MPa (**Fig. 2c**). The increasing fraction of radiatively decaying charge carriers is a result of an increasingly direct nature of the bandgap. The cross-section for vertical electron-hole recombination is increased as Rashba splitting is decreased. This model is consistent with the strong increase in PLQE with excitation density[42] as higher excitation density saturates the indirect transition and enhances the direct transition (numerically shown in literature[28]). Similar effects can also be seen in the low-temperature orthorhombic phase of MAPI where an ordered anti-ferroelectric structure is formed that suppresses internal electric fields[35]. We therefore propose that the enhancement of PLQE at high pressure has the same microscopic origin as the recently reported low temperature optoelectronic behavior.

We can exclude trap states as a major factor that causes the effects under pressure as trap states cannot account for the simultaneous increase in the PLQE and in the side peak contribution of the PL spectra. If the trap states were to become more relevant under pressure one would expect the PLQE to decrease instead.

To exclude excitonic effects as the main cause for the changes under pressure we calculated the change of exciton binding energy over different pressures (see **Supplementary Information S6** and **Fig. S5** for details). Results show that the change in exciton binding energy is less than 25% over the entire pressure range, which is not likely to result in a 5-fold decrease in charge carrier lifetime and a 2-fold increase in PLQE.

**Conclusions**

In conclusion, our results show that the prototype hybrid perovskite MAPI has an indirect bandgap. This arises due to distortion of the lead iodide framework, which leads to Rashba splitting of the conduction band. The position of the indirect transition 60 meV below the direct transition is consistent with an absorption spectrum that almost resembles a direct bandgap semiconductor. At the same time, charges thermalized in the conduction band are protected from recombination because recombination requires a change in crystal momentum. Hence, the slightly indirect bandgap explains the apparent contradiction between the long charge carrier lifetime and diffusion length, and the efficient absorption of perovskite semiconductors. Under pressure, the lattice distortion is reduced, increasing the relative strength of the direct transition, which reduces the charge carrier lifetime and increases the radiative efficiency. This tunability opens the path towards new, rational material and structural developments of perovskite semiconductors with engineered band structure. These materials could be even better suited for solar cells with a higher photovoltage, highly efficient LEDs, and lasers with low lasing threshold.

**Experimental Methods**

Methylammonium lead iodide (MAPI) thin films were prepared via spincoating in combination with anti-solvent precipitation[43]. All chemicals were purchased from Sigma-Aldrich and used as received. For the preparation of the solution, 1.2 g of lead iodide ($PbI_2$) was first mixed with 2 mL of anhydrous N,N-dimethylformamide (DMF) and heated to 100 °C while stirring. After $PbI_2$ was fully dissolved, 0.4 g of methylammonium iodide (MAI) was added to the solution. The solution was kept stirring at 100 °C until MAI was completely dissolved. A clear, yellow solution was obtained. Fused silica glass substrates were sonicated with detergent Micro 90, deionized water, acetone, and isopropanol sequentially for 10 min and oxygen plasma treated at 100 W for 15 min. For the formation of thin films, 100 µL of solution was deposited onto the glass substrates by spin-coating at 5000 rpm for 20 seconds. After 5 seconds of spinning, 100 µL of chlorobenzene was placed on the sample for the anti-solvent treatment. The sample was then transferred to a hotplate and annealed at 100 °C for 10 min. A dark brown film was obtained. A SEM image of the film can be found in Fig. S6.

Hydrostatic pressure was generated through a pressurizing liquid Fluorinert FC-72 (3M) inside a high pressure cell (ISS Inc.) using a manual pump. The pressure was first applied from ambient pressure to 400 MPa (pressure upstroke) and then down from 400 MPa to ambient pressure (pressure downstroke), both in steps of 25 MPa. A 5-7 min waiting step after application of pressure and before the measurement was chosen to allow the material to equilibrate under pressure. The pressure liquid FC-72 started to scatter a fraction of light (~5%) from 300 MPa onwards (flat spectral response, see inset Fig. S7), which we corrected for in the absorbance spectra (see below). The PL spectra and lifetimes were unaffected. From the measurement-to-measurement variability we estimate the upper limit on the error of the pressure reading (following the 5-7 min waiting step) to be 20 MPa.

Absorbance spectra of MAPI thin films were measured with a LAMBDA 750 UV/Vis/NIR Spectrophotometer (Perkin Elmer) from 550 nm to 850 nm. A correction of the absorption spectra was done for all the spectra obtained above 300 MPa by subtracting the background signal from scattering in the liquid. The background was obtained from a fit in the region of 820 nm to 850 nm where MAPI does not absorb.

Steady-state photoluminescence (PL) was measured with a home-built setup equipped with a 640 nm continuous-wave laser as source of excitation (PicoQuant LDH-D-C-640) at a power output of 0.7 mW. The PL spectra were collected at an angle of 45°. Two Thorlabs FEL-700 highpass filters were used to remove the excitation laser from the signal. The PL was coupled in a fiber connected to a OceanOptics USB4000 spectrometer, set to an integration time of 2000 ms for each measurement. The PL spectra were fitted with either one or two Voigt profiles (see Supplementary Information S2).

Time-correlated single photon counting (TCSPC) measurements were performed with a home-built setup equipped with PicoQuant PDL 828 "Sepia II" and a PicoQuant HydraHarp 400 multichannel picosecond event timer and TCSPC module. A 640 nm pulsed laser (PicoQuant LDH-D-C-640) with a repetition rate of 5 MHz was used as source of excitation and a single-photon avalanche diode (SPAD) detector (Micro Photon Devices, MPD-5CTD) was used for the detection of photoemission. A Thorlabs FEL-700 long-pass filter was used to remove the excitation laser. The TCSPC data was collected over the course of 60 seconds per measurement.

First-principles materials modelling was employed to assess the bulk response of the electronic structure to external pressure. Calculations were performed within the Kohn-Sham Density Functional Theory (DFT) formalism as implemented in VASP[44]. Projector-augmented-wave core potentials were combined with a kinetic energy cutoff of 600 eV for the plane wave basis set describing the valence electrons including spin-orbit coupling. A scalar-relativistic description of Pb[Xe] was used with the 5d semi-core electrons included as valence. The phonon-stable crystal structures reported previously[45] were taken as the starting point for a series of isobaric calculations with a stress tensor ranging from 0 to 500 MPa. A positive bandgap deformation was found (bandgap decrease with increasing pressure) in agreement with initial calculations[31]. No attempt was made to model the phase change above 325 MPa, which would require sampling changes in spatial and temporal structural disorder.

## Notes and references

1. A. Kojima, K. Teshima, Y. Shirai and T. Miyasaka, *Journal of the American Chemical Society*, 2009, **131**, 6050-6051.

2. L. Etgar, P. Gao, Z. S. Xue, Q. Peng, A. K. Chandiran, B. Liu, M. K. Nazeeruddin and M. Gratzel, *Journal of the American Chemical Society*, 2012, **134**, 17396-17399.

3. H. S. Kim, J. W. Lee, N. Yantara, P. P. Boix, S. A. Kulkarni, S. Mhaisalkar, M. Gratzel and N. G. Park, *Nano Letters*, 2013, **13**, 2412-2417.

4. J. H. Im, I. H. Jang, N. Pellet, M. Gratzel and N. G. Park, *Nature Nanotechnology*, 2014, **9**, 927-932.

5. G. C. Xing, N. Mathews, S. S. Lim, N. Yantara, X. F. Liu, D. Sabba, M. Gratzel, S. Mhaisalkar and T. C. Sum, *Nature Materials*, 2014, **13**, 476-480.

6. S. Yakunin, M. Sytnyk, D. Kriegner, S. Shrestha, M. Richter, G. J. Matt, H. Azimi, C. J. Brabec, J. Stangl, M. V. Kovalenko and W. Heiss, *Nature Photonics*, 2015, **9**, 444-U444.

7. D. Shi, V. Adinolfi, R. Comin, M. J. Yuan, E. Alarousu, A. Buin, Y. Chen, S. Hoogland, A. Rothenberger, K. Katsiev, Y. Losovyj, X. Zhang, P. A. Dowben, O. F. Mohammed, E. H. Sargent and O. M. Bakr, *Science*, 2015, **347**, 519-522.

8. Y. Bi, E. M. Hutter, Y. J. Fang, Q. F. Dong, J. S. Huang and T. J. Savenije, *Journal of Physical Chemistry Letters*, 2016, **7**, 923-928.

9. Q. F. Dong, Y. J. Fang, Y. C. Shao, P. Mulligan, J. Qiu, L. Cao and J. S. Huang, *Science*, 2015, **347**, 967-970.

10. H. S. Kim, C. R. Lee, J. H. Im, K. B. Lee, T. Moehl, A. Marchioro, S. J. Moon, R. Humphry-Baker, J. H. Yum, J. E. Moser, M. Gratzel and N. G. Park, *Scientific Reports*, 2012, **2**.

11. T. Baikie, Y. N. Fang, J. M. Kadro, M. Schreyer, F. X. Wei, S. G. Mhaisalkar, M. Graetzel and T. J. White, *Journal of Materials Chemistry A*, 2013, **1**, 5628-5641.

12. A. Cuevas and D. Macdonald, *Solar Energy*, 2004, **76**, 255-262.

13. D. W. deQuilettes, S. M. Vorpahl, S. D. Stranks, H. Nagaoka, G. E. Eperon, M. E. Ziffer, H. J. Snaith and D. S. Ginger, *Science*, 2015, **348**, 683-686.

14. X. Y. Zhu and V. Podzorov, *Journal of Physical Chemistry Letters*, 2015, **6**, 4758-4761.

15. H. H. Fang, R. Raissa, M. Abdu-Aguye, S. Adjokatse, G. R. Blake, J. Even and M. A. Loi, *Advanced Functional Materials*, 2015, **25**, 2378-2385.

16. C. Motta, F. El-Mellouhi, S. Kais, N. Tabet, F. Alharbi and S. Sanvito, *Nature Communications*, 2015, **6**.

17. F. Brivio, K. T. Butler, A. Walsh and M. van Schilfgaarde, *Physical Review B*, 2014, **89**.

18. T. Etienne, E. Mosconi and F. De Angelis, *Journal of Physical Chemistry Letters*, 2016, **7**, 1638-1645.

19. M. Kepenekian, R. Robles, C. Katan, D. Sapori, L. Pedesseau and J. Even, *ACS Nano*, 2015, **9**, 11557-11567.

20. F. Zheng, L. Z. Tan, S. Liu and A. M. Rappe, *Nano Letters*, 2015, **15**, 7794-7800.

21. D. Niesner, M. Wilhelm, I. Levchuk, A. Osvet, S. Shrestha, M. Batentschuk, C. Brabec and T. Fauster, *Physical Review Letters*, 2016, **117**, 126401.

22. F. J. Manjon, A. Segura, V. Munoz-Sanjose, G. Tobias, P. Ordejon and E. Canadell, *Physical Review B*, 2004, **70**.

23 D. Errandonea, E. Bandiello, A. Segura, J. J. Hamlin, M. B. Maple, P. Rodriguez-Hernandez and A. Munoz, *Journal of Alloys and Compounds*, 2014, **587**, 14-20.

24 T. J. Ou, J. J. Yan, C. H. Xiao, W. S. Shen, C. L. Liu, X. Z. Liu, Y. H. Han, Y. Z. Ma and C. X. Gao, *Nanoscale*, 2016, **8**, 11426-11431.

25 A. Jaffe, Y. Lin, C. M. Beavers, J. Voss, W. L. Mao and H. I. Karunadasa, *ACS Central Science*, 2016, **2**, 201-209.

26 S. J. Jiang, Y. A. Fang, R. P. Li, H. Xiao, J. Crowley, C. Y. Wang, T. J. White, W. A. Goddard, Z. W. Wang, T. Baikie and J. Y. Fang, *Angewandte Chemie-International Edition*, 2016, **55**, 6540-6544.

27 L. Kong, G. Liu, J. Gong, Q. Hu, R. D. Schaller, P. Dera, D. Zhang, Z. Liu, W. Yang, K. Zhu, Y. Tang, C. Wang, S.-H. Wei, T. Xu and H.-k. Mao, *Proceedings of the National Academy of Sciences*, 2016, **113**, 8910-8915.

28 P. Azarhoosh, S. McKechnie, J. M. Frost, A. Walsh and M. van Schilfgaarde, *APL Materials*, 2016, **4**, 091501.

29 J. Feng, *APL Materials*, 2014, **2**.

30 J. Tauc, *Materials Research Bulletin*, 1968, **3**, 37-46.

31 S. H. Wei and A. Zunger, *Physical Review B*, 1997, **55**, 13605-13610.

32 J. M. Frost, K. T. Butler, F. Brivio, C. H. Hendon, M. van Schilfgaarde and A. Walsh, *Nano Letters*, 2014, **14**, 2584-2590.

33 R. L. Milot, G. E. Eperon, H. J. Snaith, M. B. Johnston and L. M. Herz, *Advanced Functional Materials*, 2015, **25**, 6218-6227.

34 K. W. Wu, A. Bera, C. Ma, Y. M. Du, Y. Yang, L. Li and T. Wu, *Physical Chemistry Chemical Physics*, 2014, **16**, 22476-22481.

35. G. Grancini, A. R. S. Kandada, J. M. Frost, A. J. Barker, M. De Bastiani, M. Gandini, S. Marras, G. Lanzani, A. Walsh and A. Petrozza, *Nature Photonics*, 2015, **9**, 695-701.
36. V. D'Innocenzo, G. Grancini, M. J. P. Alcocer, A. R. S. Kandada, S. D. Stranks, M. M. Lee, G. Lanzani, H. J. Snaith and A. Petrozza, *Nature Communications*, 2014, **5**.
37. C. Wehrenfennig, M. Z. Liu, H. J. Snaith, M. B. Johnston and L. M. Herz, *APL Materials*, 2014, **2**.
38. X. X. Wu, M. T. Trinh, D. Niesner, H. M. Zhu, Z. Norman, J. S. Owen, O. Yaffe, B. J. Kudisch and X. Y. Zhu, *Journal of the American Chemical Society*, 2015, **137**, 2089-2096.
39. A. Sadhanala, F. Deschler, T. H. Thomas, S. E. Dutton, K. C. Goedel, F. C. Hanusch, M. L. Lai, U. Steiner, T. Bein, P. Docampo, D. Cahen and R. H. Friend, *Journal of Physical Chemistry Letters*, 2014, **5**, 2501-2505.
40. M. Saba, M. Cadelano, D. Marongiu, F. P. Chen, V. Sarritzu, N. Sestu, C. Figus, M. Aresti, R. Piras, A. G. Lehmann, C. Cannas, A. Musinu, F. Quochi, A. Mura and G. Bongiovanni, *Nature Communications*, 2014, **5**.
41. W. G. Kong, Z. Y. Ye, Z. Qi, B. P. Zhang, M. Wang, A. Rahimi-Iman and H. Z. Wu, *Physical Chemistry Chemical Physics*, 2015, **17**, 16405-16411.
42. F. Deschler, M. Price, S. Pathak, L. E. Klintberg, D. D. Jarausch, R. Higler, S. Huttner, T. Leijtens, S. D. Stranks, H. J. Snaith, M. Atature, R. T. Phillips and R. H. Friend, *Journal of Physical Chemistry Letters*, 2014, **5**, 1421-1426.
43. M. L. Petrus, T. Bein, T. J. Dingemans and P. Docampo, *Journal of Materials Chemistry A*, 2015, **3**, 16874-16874.
44. G. Kresse and D. Joubert, *Physical Review B*, 1999, **59**, 1758-1775.

45 F. Brivio, J. M. Frost, J. M. Skelton, A. J. Jackson, O. J. Weber, M. T. Weller, A. R. Goni, A. M. A. Leguy, P. R. F. Barnes and A. Walsh, *Physical Review B*, 2015, **92**.

Supplementary Information for:

# Indirect to direct bandgap transition in methylammonium lead halide perovskite

**Authors:** Tianyi Wang[1†], Benjamin Daiber[1†], Jarvist M. Frost[2], Sander A. Mann[1], Erik C. Garnett[1], Aron Walsh[2], Bruno Ehrler[1]*

**Affiliations:**

[1] Center for Nanophotonics, FOM Institute AMOLF, Science Park 104, 1098 XG Amsterdam, The Netherlands.

[2] Centre for Sustainable Chemical Technologies and Department of Chemistry, University of Bath, Claverton Down, Bath BA2 7AY, United Kingdom.

*Correspondence to: ehrler@amolf.nl

†These authors contributed equally to the work.

**Table of content**

## Methods

### S1. Analysis of absorption spectra

Tauc plots are used to track the bandgap shifts of MAPI under pressure. Here we use the Tauc plots for direct bandgap semiconductors to fit the absorption spectrum as the direct transition is dominating the absorption for MAPI due to the high density of state in the valence band[S1]. The Tauc equation for direct bandgap semiconductors is written as:

$$\alpha h\nu = A(h\nu - E_G)^{1/2}$$

where $\alpha$ is the absorption coefficient, $h\nu$ is the photon energy, $A$ is a constant and $E_G$ is the bandgap. This analysis requires parabolic bands, which is a valid assumption for MAPI only very close to the (direct transition) absorption onset[S2].

The absorption coefficient is calculated with the following equation:

$$\alpha = \left(\frac{2.303}{t}\right) * OD$$

Where $t$ is the thickness of the film (400 nm), 2.303 is the conversion constant between log(x) and ln(x) and $OD$ is the optical density (absorbance measured with UV/Vis spectroscopy).

We plot $(\alpha h\nu)^2$ against photon energy. A straight line is fitted by minimizing the root mean square difference, to the linear region of the Tauc plots. Extrapolation of this straight line to the x-axis (photon energy) is taken as the bandgap. The quality of the fitting procedure is evaluated by a reduced chi squared metric. Fits with reduced chi squared between 0.95 – 1.00 are considered to be "good" fits. Fits to the linear region of the Tauc spectrum at all pressures fulfill this criteria. We define our error in the bandgap estimation as the standard error of the linear fit.

## S2. Fitting of steady-state PL spectra

The PL spectra are best fitted with Voigt profiles[S3], which include the contributions from both Gaussian and Lorentzian components. The Voigt profile is given by

$$P(x,y) = \frac{1}{w_G}\sqrt{\frac{ln2}{\pi}} K(x,y),$$

where $K(x,y)$, the so-called "Voigt function", is given by

$$K(x,y) = \frac{y}{\pi}\int_{-\infty}^{\infty}\frac{exp(-t^2)}{y^2 + (x-t)^2}dt$$

with $x = \frac{v-v_0}{w_G}\sqrt{ln2}$, $y = \frac{w_L}{w_G}\sqrt{ln2}$, and where $w_G$ and $w_L$ are the half widths of the Gaussian and Lorentzian components respectively, $v$ is the wave number, and $v_0$ is the wave number at the line center.

The goodness of fit is evaluated by adjusted R squared (also known as coefficient of determination). Fits with adjusted R squared between 0.95 – 1.00 are considered to be "good" fits and all the Voigt fits for PL spectra fulfill this criterion. We define our error in the peak position as the standard errors of the Voigt fit.

The Voigt profile shape is chosen because it (a) offers the best description of the data since the emission broadening is a combination of homogeneous (Lorentzian) and inhomogeneous (Gaussian) broadening[S4, S5] and (b) is the most general curve shape, as it is a convolution of Gaussian and Lorentzian line shapes.

## S3. Estimation of Phonon energy

The phonon momentum can be calculated by assuming two parabolic bands that have a 60 meV difference from the k=0 point and the band edge. The curvature of the parabolic bands is determined by the effective masses, taken from literature[S6]. The resulting wave vector is k= 0.048 Å$^{-1}$. This value is comparable to k~0.05 Å$^{-1}$ from literature[S7]. To calculate the energy of an acoustic phonon with this wave vector we use the phonon dispersion relation[S8];

$$\omega = 2\sqrt{\gamma/M}\,|sin(k\,a/2)|$$

where $\omega$ is the angular frequency, $\gamma$ is the force constant, $M$ is the mass of the unit cell, $k$ is the phonon wave vector, and $a$ is the lattice constant.

The force constant $\gamma$ is connected to Young's modulus $Y$ and the lattice parameter $a$ via $\gamma = Y \times a$. The Young's modulus was derived from the phonon dispersion relation in literature[S9] and found to be Y=(13 ± 2) GPa. The lattice parameter is a=6.4 Å[S10]. The mass of the unit cell is the sum of all unit cell constituents. These values lead to a phonon energy of 0.6 meV, for a phonon with momentum such that the indirect transition can occur.

## S4. Fit of photothermal deflection spectroscopy (PDS) data

To fit the photothermal deflection spectroscopy data, we use equations for the direct bandgap, the indirect bandgap, and the Urbach tail. We extracted the absorption data for MAPI from[S11] (Sadhanala et al.) using the website (http://arohatgi.info/WebPlotDigitizer/) (**Fig. S8**). As above, the data was then linearized according to the Tauc rule for both direct and indirect bandgaps:

$$\alpha h\nu = A(h\nu - E_G)^r$$

with the absorption coefficient $\alpha$, photon energy $h\nu$, the scaling constant $A$, the bandgap energy $E_G$ and the coefficient $r$ ($r = 2$ for indirect allowed transition, $r = ½$ for direct allowed transition)[S12, S13]. Then the linear region of the plot is identified and a straight line is fitted (with Mathematica 10.3.1 using LinearModelFit) (**Fig. S8**). The beginning of the absorbance can be fitted separately with an exponential, the Urbach tail, usually assigned to shallow traps and Gaussian disorder[S14]. The Urbach energy we extract is 21 meV, comparable to the 15 meV measured before[S11, S15].

The dashed lines depict an attempt of fitting the whole bandedge with only a direct bandedge and an Urbach tail.

The fit with only a direct bandgap does not describe the data well, with non-Gaussian distributed residuals, which led us to the conclusion that an indirect bandgap is also supported from PDS absorption data.

Additional PDS data is also reported by Zhang et al.[S16] (**Fig. S9**) and de Wolf et al.[S15] (**Fig. S10**). Both datasets are also consistent with an indirect bandgap, with a more dominant Urbach tail in the de Wolf et al. data (**Fig. S10**).

**S5. Analysis of TCSPC data**

First the differential equation relating the decay of charge carrier density and the bimolecular (assumed radiative) and monomolecular (assumed non-radiative) decay is solved (see **Equation 1** of the main text). This model does not include Auger recombination, since this contribution is negligible, as shown below. This model assumes that an equal amount of electrons and holes are generated upon photoexcitation, and that the excitation leads to a much higher charge carrier density than the intrinsically available charge carrier density. Both electrons and holes are treated as equivalent charge carriers since in MAPI perovskites mobility of holes and electrons is comparable[S17]. We normalize the data at $t = 0$ and set the boundary condition for the charge carrier density in the fit accordingly to

$$n(t = 0) = n^0 = 1/cm^3.$$

An alternative approach would be to use $n(t=0)=n^0$ as a fit parameter. However, this leads to the problem that $n^0$ and $k_R$ are only weakly independent and therefore cannot be fitted independently. The fit then shows varying $n^0$ at different pressure values, which is unphysical.

We assume a consistent $n^0$ across all pressures. This implies that the initial charge carrier density is constant across different pressures, assuming that the laser fluence and the absorption are constant. We confirmed that the laser power at the sample was constant (within <5 %, see below), therefore the number of photo-induced charge carriers should not change significantly either. The pressure liquid (Fluorinert FC-72, 3M) becomes slightly translucent at around 300 MPa which could lead to a decreased laser power at the sample, therefore to a decrease in charge carrier density. However, by placing a silicon photodiode inside the pressure cell we found that the power variation under pressure is smaller than 5% across the entire range from 0 MPa to 400 MPa. We assume that the absorption efficiency of the excitation laser by MAPI is independent of pressure, as the change to the physical structure and therefore electronic structure is a small perturbation, until the phase transition to the orthorhombic form.

The solution to Equation 1 in the main text describes the charge carrier decay as a function of time:

$$n(t) = \frac{k_{NR}n^0}{e^{k_{NR}t}(k_{NR}+n^0 k_R)-n^0 k_R}$$

(Units: $k_{Nr} \to \frac{1}{s}$, $k_R \to \frac{cm^3}{s}$), $n(t=0) = n^0 = 1/c\,m^3$. Photoluminescence (PL) requires radiative recombination of two charge carriers, therefore the PL signal is proportional to $n^2$. This is based on the assumption that monomolecular recombination is (mostly) non-radiative. Since our data is well fitted by this model, and it is also reported for MAPI that dark trap states are responsible for monomolecular recombination [S18], this assumption is valid:

$$Pl(t) = An(t)^2 + background$$

The proportionality factor *A* includes the number of generated photons (dependent on the charge carrier density $n^0$) and the outcoupling efficiency. The sum of *A* and the background is described by the height of the PL decay curve at *t = 0* in our model with normalized $n^0$. The background is determined independently by fitting the average of the datapoints before the excitation pulse. The Instrument Response Function (IRF) at 640 nm has a FWHM of 1 ns, so to exclude any effects of the IRF on our analysis we start fitting from 1 ns onwards. Fitting uses the NonLinearFit function in Mathematica 10.3. This algorithm minimizes the Chi-squared value to find the best fit while using 1/cts as weighing factor since $\sqrt{cts}$ is the error on (Poisson-distributed) counting data. The Chi-squared method assumes that the datapoints are scattered around the "real" curve following a normal distribution. As we are *counting* photons in photoluminescence measurements, the real scatter histogram follows a Poisson distribution. If the mean of a Poisson distribution is sufficiently high (conventionally μ > 10) it can be approximated by a normal distribution. We implement this by fitting the data only until the intensity has reached background + 15 counts. In this range the residuals should follow the normal distribution in good approximation. This null hypothesis is then tested with a Cramér-von Mises test, resulting in a reliable goodness of fit measure for our highly nonlinear model (Chi-Squared reduced close to 1 should not be used with nonlinear models as shown in literature[S19]. The p-values of the Cramer-Von Mises test are shown in **Fig. S11**, all but three have a significance level $P > 0.05$. Residuals of the three fits with $P < 0.05$ are included in **Fig. S11** and show no structure but a few outliers producing the low P-Values, hence the fit can also be accepted in those cases. The results of the fitted parameters can be found in the main text.

We define the radiative efficiency as follows:

$$\Theta_{Rad} = \frac{chargecarriersdecayedbimolecularly}{alldecayedchargecarriers} = \frac{\int_{t=0}^{\infty} k_{bi} n(t)^2 dt}{\int_{t=0}^{\infty} [k_{bi} n(t)^2 + k_{mono} n(t)] dt}$$

Similarly, the fraction of charge carriers decayed non-radiatively is defined as

$$\Theta_{NonRad} = \frac{\int_{t=0}^{\infty} k_{mono} n(t) dt}{\int_{t=0}^{\infty} [k_{bi} n(t)^2 + k_{bi} n(t)] dt}$$

while $\Theta_{NonRad} + \Theta_{Rad} = 1$. Both values are plotted in **Fig. 2C** of the main text.

***On Auger recombination:*** To determine the influence of Auger recombination ($\propto n^3$) and to show that the two-process model used here is sufficient, we carried out TCSPC measurements at different excitation densities. At 0 MPa, we varied the laser power by one order of magnitude compared to the other measurements. Since Auger recombination is a fast process it would be visible in the onset of the decay, rising with the cube of the excitation density. As seen in **Fig. S4**, no evidence is shown for Auger recombination in our TCSPC measurements.

***Influence of charge carrier density:*** The increased radiative efficiency we observe could also arise through higher charge carrier densities, as discussed in literature[S20, S21] and from the dependence of $\Theta_{Rad}$ which increases with charge carrier density. The laser power was held constant (variation <5%, photodiode measurement, see above) during the measurement and the sample was fixed in a sample holder, restricting movement and therefore fixing the incoming laser power density. To show that even large variations in charge carrier densities do not lead to our observed simultaneous increase in $k_{mono}$ and $k_{bi}$ we fitted the TCSPC data where we changed the laser intensity with the model assuming constant laser power. This shows an increase in $k_{bi}$ (**Fig. S12**) but a stable $k_{mono}$ (**Fig. S13**) with increasing laser power, different from the trend in the pressure TCSPC experiment. The apparent increase in $k_{bi}$ extracted from the fit is due to the fast radiative component becoming more important with higher charge carrier density and the model can only increase the value of $k_{bi}$ to optimize the fit when the amplitude is fixed. A decrease in charge carrier density over pressure would lead to a decrease in radiative efficiency, opposite to the trend of increasing radiative efficiency in the pressure measurement. If we take the model with varying initial charge carrier density and fit the data with $n_0$ as the only fitting parameter, (holding $k_{mono}$ and $k_{bi}$ constant), we are able to fit the data taken at different excitation densities (**Fig. S14**).

### *Absorbed charge carrier density:*

To calculate the charge carrier density we place a power meter at the same position as our film inside the pressure cell and extract a total power of 37μW. To calculate the power density, we replace the power meter with a beam profiler (Thorlabs BC106N-VIS/M, **Fig. S15**). We measure a peak power density of 62 mW/cm$^2$, at which we observe a radiative efficiency of 20 % at ambient pressure. This is in agreement with a PLQE of <40 % at fluences of around 50 mW cm$^{-2}$ reported in literature[S21].

The repetition rate of excitation is 5 MHz, resulting in a excitation energy of 12.4 nJ/cm$^2$ per pulse. To calculate the charge carrier density we use:

$$n_{initial} = \frac{PowerDensity * Absorbance(640nm)}{E_{photon}(640nm) * d_{sample}} * geometry factor$$

With the measured power density, absorbance of MAPI at 640 nm, the energy of a photon at 640 nm, the thickness of the sample $d_{sample} = 400\ nm$ and a geometry factor, which accounts for the additional refraction because we measured the beam profile in air and not in liquid as in the pressure experiment, and the fact that we measured at 45° sample to beam orientation. For the calculation of the geometry factor the beam is first propagated until it hits the silica window where the size can be calculated from the distance and the focal length of the lens. The beam is then refracted twice, once at the air-silica and then at the silica-liquid interface. The beam is then propagated to the sample, where the beam radius is calculated again. This results in a ratio of the squares of these radii of 2.14. In addition, the sample is tilted by 45° from the incoming beam during the pressure measurement, so we add a factor of cos(45°) resulting in a total geometry factor of 1.51.

This leads to an initial charge carrier density of $6.05 \times 10^{14}\ cm^{-3}$.

***Comparison with literature values for the rate constants:*** Wehrenfennig et al. carried out a measurement of the rate constants at ambient pressure on films of the same material using THz techniques[S21]. We compare this with our measurement at 0 MPa. Wehrenfennig et al. measured at charge carrier densities of $10^{17}$-$10^{19}\ cm^{-3}$, considerably higher than our density of $10^{14}\ cm^{-3}$. They report a non-radiative recombination rate of 14 μs$^{-1}$, whereas we measure (20.0 ± 0.2) μs$^{-1}$, with the error derived from the fit. They report a radiative recombination rate of $k_{bi}^{lit.} = 9.2 \times 10^{-10}\ cm^3\ s^{-1}$. Our calculation is using an initial charge carrier density of unity (1 cm$^{-3}$). Therefore $k_{bi}^{Real} = \frac{k_{bi}^{fit}}{n(t=0)}$, here $(1.85 \pm 0.05) \times 10^{-8}\ cm^3\ s^{-1}$, around 20 times higher than reported by Wehrenfennig et al. The vastly different excitation densities[S7] as well as the different techniques

(TCSPC vs. THz photoconductivity measurements) and batch-to-batch variations might explain the difference.

**S6. Exciton binding energy under pressure**

The influence of excitonic behavior is investigated as a possible explanation of the PL side peak. One could conceive that under pressure, restricted movement of the methylammonium ions could lead to a change in exciton binding energy, leading to the changes in optoelectronic behavior we observe. The absorbance data (**Fig. 1A**) is analyzed for a change in excitonic peak height using the Elliott formula [S22]. We see no change in exciton binding energy within the error of the measurement (26% relative error) (**Fig. S5**). The energy difference between main and side peak of ~60 meV also stands in contrast to reported exciton binding energies of ~10 meV, now generally accepted in the field[S21].

Absorption data can be described by contributions of an exitonic peak and a band continuum [S23].

$$\alpha(E) \propto \mu_{cv}^2 \sqrt{E_B} \left( \sum_n \alpha_{nx} + \alpha_c \right)$$

where $\mu_{cv}^2$ is the transition dipole moment, $E_B$ is the exciton binding energy, $\alpha_{nx}$ is the absorption from the n-th exciton, and $\alpha_c$ is the continuum absorption. We consider the ratio of absorbance at the excitonic peak of the i-th pressure measurement ($\alpha_i$) with the ambient pressure measurement ($\alpha_0$):

$$\frac{\alpha_i}{\alpha_0} = \frac{\mu_{cv}^2 \sqrt{E_{Bi}} (\sum_n \alpha_{nx} + \alpha_c)}{\mu_{cv}^2 \sqrt{E_{B0}} (\sum_n \alpha_{nx} + \alpha_c)} = \frac{\sqrt{E_{Bi}} (\sum_n \alpha_{nx} + \alpha_c)}{\sqrt{E_{B0}} (\sum_n \alpha_{nx} + \alpha_c)}$$

To calculate an upper bound of change in exciton binding energy we assume a strong continuum contribution, and consider the limit of $\alpha_c \to \infty$. This limit leads to:

$$\frac{\alpha_i}{\alpha_0} = \frac{\sqrt{E_{Bi}}}{\sqrt{E_{B0}}} \Rightarrow \frac{E_{Bi}}{E_{B0}} = \left(\frac{\alpha_i}{\alpha_0}\right)^2$$

To get the value of $\frac{\alpha_i}{\alpha_0}$, we determine the maxima of the excitonic peak. To show the excitonic peak more clearly than in **Fig. 1A** we treat the data. We fit a linear equation to the low wavelength realm in the data of **Fig. 1A** for every curve and subtract the respective line from the data (**Fig. S5**)

We use the function Lowpass in Mathematica 10.3 to filter the noise and extract the maximum peak value for each pressure point seen in **Fig. S5**. We then calculate the change in exciton binding energy from the peak absorbance data points (**Fig. S5**). The upper bound for the increase in exciton binding energy is 26% with a large standard deviation of 11% also represented in the large error bars. Error bars are taken from the standard deviation of the noise, propagated via a Gaussian error propagation according to the formula above.

**References**

S1.     J. Tauc, *Materials Research Bulletin*, 1968, **3**, 37-46.

S2.     F. Brivio, K. T. Butler, A. Walsh and M. van Schilfgaarde, *Physical Review B*, 2014, **89**.

S3.     B. H. Armstrong, *Journal of Quantitative Spectroscopy and Radiative Transfer*, 1967, **7**, 61-88.

S4.     A. D. Wright, C. Verdi, R. L. Milot, G. E. Eperon, M. A. Pérez-Osorio, H. J. Snaith, F. Giustino, M. B. Johnston and L. M. Herz, *Nature Communications*, 2016, **7**, 0.

S5.     C. Wehrenfennig, M. Liu, H. J. Snaith, M. B. Johnston and L. M. Herz, *The Journal of Physical Chemistry Letters*, 2014, **5**, 1300-1306.

S6.     J. M. Frost, K. T. Butler, F. Brivio, C. H. Hendon, M. van Schilfgaarde and A. Walsh, *Nano Letters*, 2014, **14**, 2584-2590.

S7.     P. Azarhoosh, S. McKechnie, J. M. Frost, A. Walsh and M. van Schilfgaarde, *APL Materials*, 2016, **4**, 091501.

S8.     C. Kittel, *Introduction to Solid State Physics*. *John Wiley and Sons, Inc., New York,* **8th edition,** 2005.

S9.     Beecher, A. N. *et al.* The cubic phase of methylammonium lead iodide perovskite is not locally cubic. 1–10 (2016).

S10.    Oku, T. *Sol. Cells - New Approaches Rev.* 2015, 77–101. doi:10.5772/58490

S11.    A. Sadhanala, F. Deschler, T. H. Thomas, S. E. Dutton, K. C. Goedel, F. C. Hanusch, M. L. Lai, U. Steiner, T. Bein, P. Docampo, D. Cahen and R. H. Friend, *Journal of Physical Chemistry Letters*, 2014, **5**, 2501-2505.

S12.    J. Tauc, R. Grigorovici and A. Vancu, *physica status solidi (b)*, 1966, **15**, 627-637.

S13.    E. A. Davis and N. F. Mott, *Philosophical Magazine*, 1970, **22**, 0903-0922.

S14.    S. John, C. Soukoulis, M. H. Cohen and E. N. Economou, *Physical Review Letters*, 1986, **57**, 1777-1780.

S15.    S. De Wolf, J. Holovsky, S. J. Moon, P. Loper, B. Niesen, M. Ledinsky, F. J. Haug, J. H. Yum and C. Ballif, *The Journal of Physical Chemistry Letters*, 2014, **5**, 1035-1039.

S16.	W. Zhang, M. Saliba, D. T. Moore, S. K. Pathak, M. T. Horantner, T. Stergiopoulos, S. D. Stranks, G. E. Eperon, J. A. Alexander-Webber, A. Abate, A. Sadhanala, S. Yao, Y. Chen, R. H. Friend, L. A. Estroff, U. Wiesner and H. J. Snaith, *Nature Communications*, 2015, **6**, 6142.

S17.	C. Motta, F. El-Mellouhi and S. Sanvito, *Scientific Reports*, 2015, **5**, 12746.

S18.	G.-J. A. H. Wetzelaer, M. Scheepers, A. M. Sempere, C. Momblona, J. Ávila and H. J. Bolink, *Advanced Materials*, 2015, **27**, 1837-1841.

S19.	R. Andrae, T. Schulze-Hartung, P. Melchior, Dos and don'ts of reduced chi-squared, 2010, arXiv:1012.3754v1

S20.	F. Deschler, M. Price, S. Pathak, L. E. Klintberg, D. D. Jarausch, R. Higler, S. Huttner, T. Leijtens, S. D. Stranks, H. J. Snaith, M. Atature, R. T. Phillips and R. H. Friend, *The Journal of Physical Chemistry Letters*, 2014, **5**, 1421-1426.

S21.	C. Wehrenfennig, G. E. Eperon, M. B. Johnston, H. J. Snaith and L. M. Herz, *Advanced Materials*, 2014, **26**, 1584-1589.

S22.	L. M. Herz, *Annual Reviews of Physical Chemistry*, 2016, **67**, 65-89.

S23.	N. Sestu, M. Cadelano, V. Sarritzu, F. Chen, D. Marongiu, R. Piras, M. Mainas, F. Quochi, M. Saba, A. Mura and G. Bongiovanni, *The Journal of Physical Chemistry Letters*, 2015, **6**, 4566-4572.

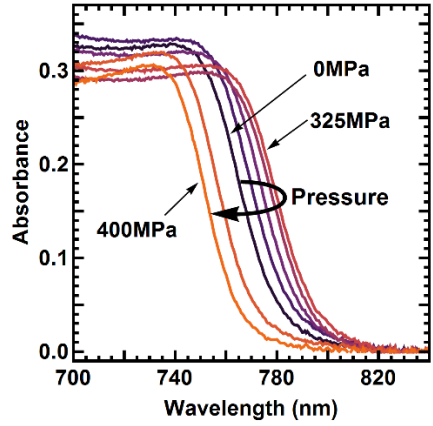

**Fig. S1.** Absorbance Spectra at different pressures, from dark to light: 0 MPa, 100 MPa, 200 MPa, 300 MPa, 325 MPa, 350 MPa, 400 MPa.

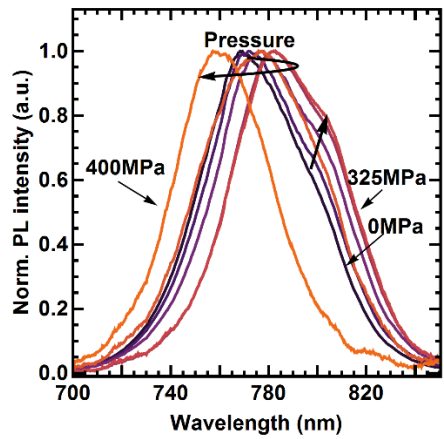

**Fig. S2.** Normalized steady-state PL spectra at different pressure, from dark to light: 0 MPa, 100 MPa, 200 MPa, 300 MPa, 325 MPa, 350 MPa, 400 MPa.

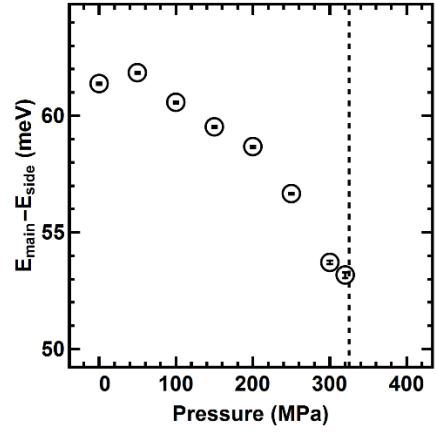

**Fig. S3.** Difference in the PL peak position between the main and the side peak.

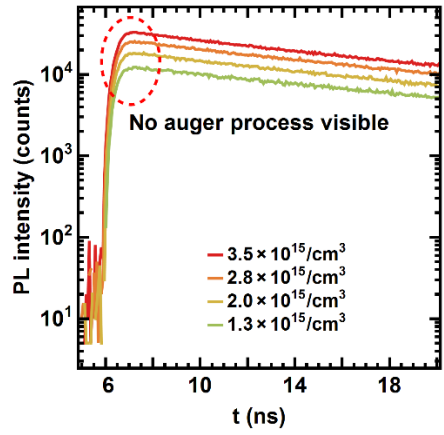

**Fig. S4.** Initial decay of the PL signal at different excitation densities at ambient pressure.

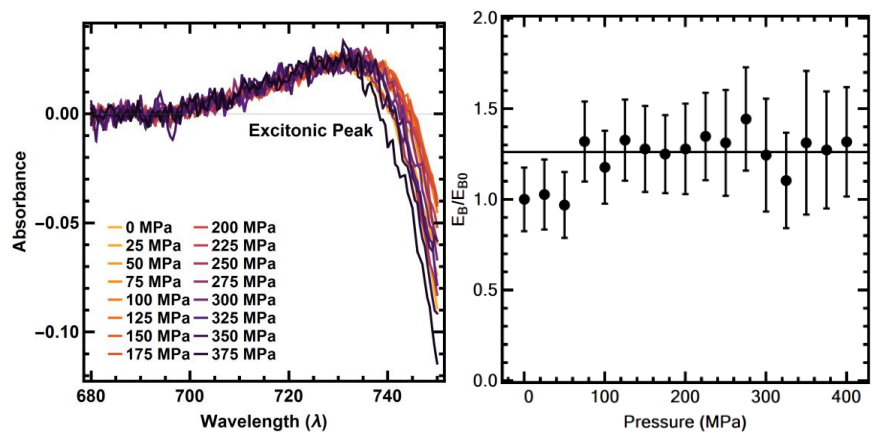

**Fig. S5 Left:** Excitonic peak in absorbance data. No change in height (indicating a change in exciton binding energy) is visible in the raw data. **Right:** Change in exciton binding energy in percent over pressure. No structure is visible, the variation is likely from measurement noise. Black line is the mean, error bars from the noise standard deviation.

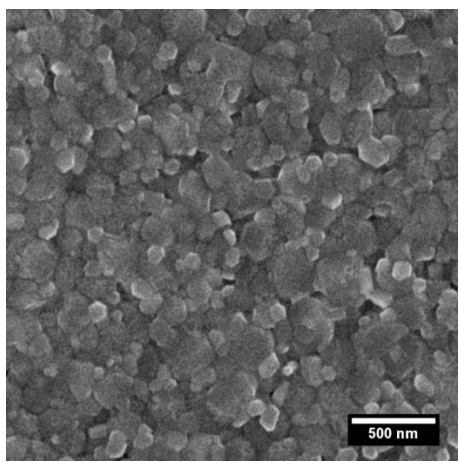

**Fig. S6.** Scanning electron micrograph of MAPI thin film on quartz substrate.

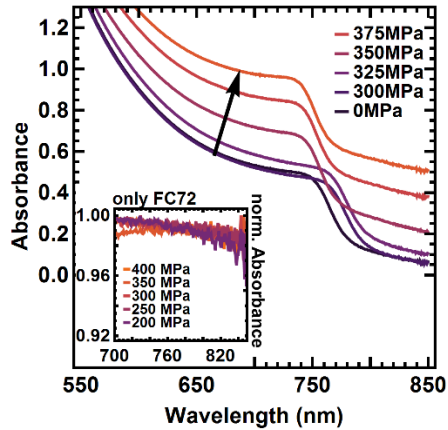

**Fig. S7.** Absorption spectra with transparent and translucent pressure liquid. The inset shows the normalized absorption spectra of the pressure liquid without sample at elevated pressure. The spectral response is flat around the band-edge.

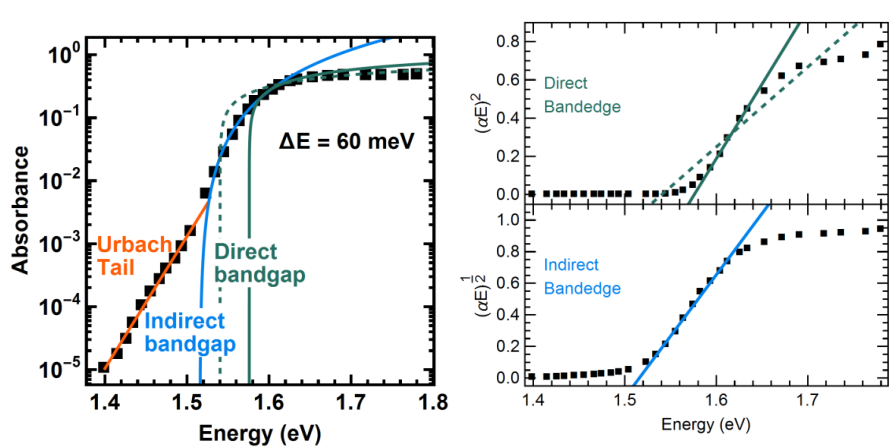

**Fig. S8 Left:** PDS data from Sadhanala et al.[S9] fit with an exponential Urbach tail (orange) and indirect (blue) and direct (green) bandgaps. A fit of only a direct bandgap and Urbach tail is shown in dashed green. The energy distance between direct and indirect bandgap is 60 meV. **Right:** Linearized data according to tauc rule with linear fits.

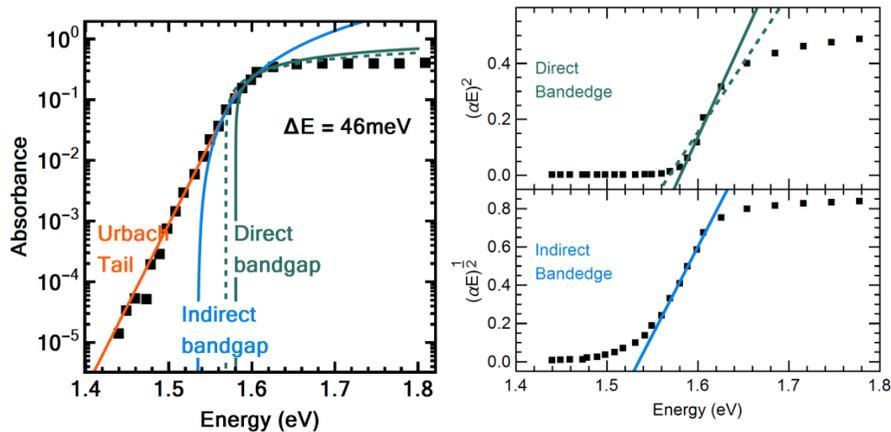

**Fig. S9 Left:** PDS data from de Wolf et al. [S13] fit with an exponential Urbach tail (orange) and indirect (blue) and direct (green) bandgaps. A fit of only a direct bandgap and Urbach tail is shown in dashed green. The energy distance between direct and indirect bandgap is 46 meV. A large Urbach tail overlaps with the indirect bandgap. **Right:** Linearized data according to tauc rule with linear fits.

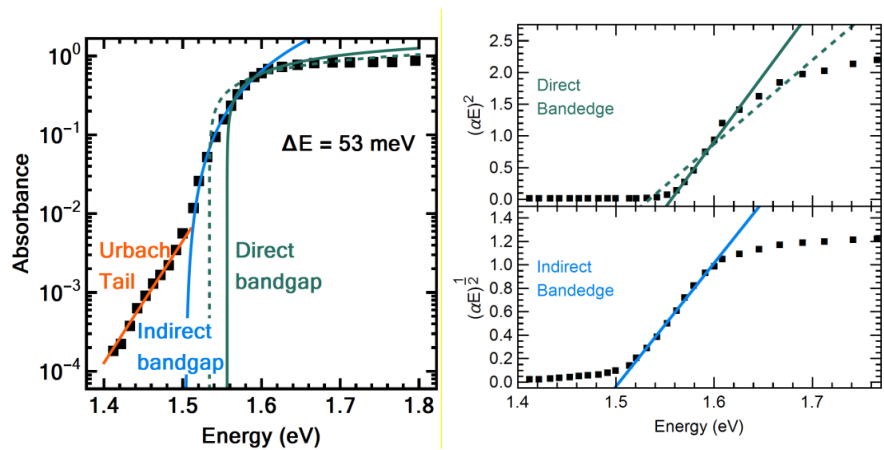

**Fig. S10 Left:** PDS data from Zhang et al. [S14] fit with an exponential Urbach tail (orange) and indirect (blue) and direct (green) bandgaps. A fit of only a direct bandgap and Urbach tail is shown in dashed green. The energy distance between direct and indirect bandgap is 53 meV. **Right:** Linearized data according to tauc rule with linear fits.

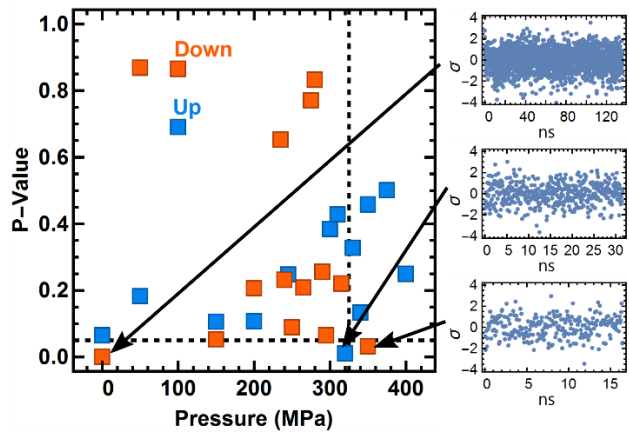

**Fig. S11. Left:** P-values of Cramer-Von Mises test, testing normality for TCSPC fitting at different pressures. Horizontal dashed line is the 5 % significance level, vertical dashed line the phase transition at 325 MPa. **Right:** Standardized residuals of the fit at values below significance level showing outliers.

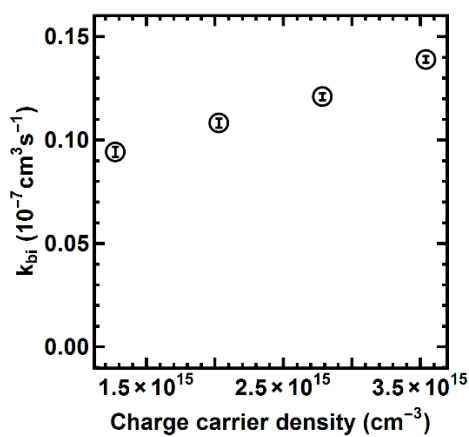

**Fig. S12.** Data seen in Fig. S4 fitted to our model forcing constant charge carrier density. Because the actual charge carrier density varies, the model has to increase the bimolecular rate to account for a faster onset of decay.

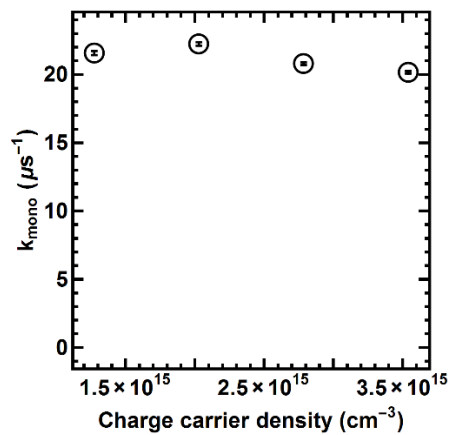

**Fig. S13.** Data seen in Fig. S4 fitted to our model forcing constant charge carrier density. The monomolecular rate stays largely unaffected by the change in charge carrier density.

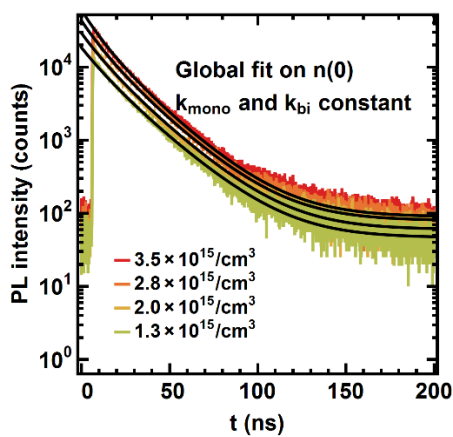

**Fig. S14.** Data shown in Fig. S4 fitted globally while allowing for varying charge carrier density. Both monomolecular and bimolecular rates are fixed to the rates retrieved under ambient pressure. The faster decay onset is reproduced correctly by an increase in charge carrier density.

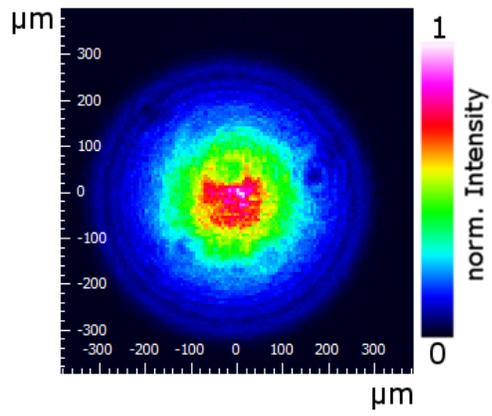

**Fig. S15.** Beam profile measured in air with the same lens and the same lens to sample distance as in the pressure experiment.